\documentclass{aastex62}

\usepackage{graphicx}
\usepackage{amssymb}
\usepackage{amsmath}
\usepackage{url}

\received{21 March 2019}
\revised{8 October 2019}
\accepted{11 October 2019}

\submitjournal{ApJ}

\shorttitle{Effect of Land Fraction \& SED on Planetary Albedo}
\shortauthors{Rushby et al.}

\begin{document}

\title{The Effect of Land Fraction and Host Star Spectral Energy Distribution\\ on the Planetary Albedo of Terrestrial Worlds}

\correspondingauthor{Andrew J. Rushby}
\email{arushby@uci.edu}

\author[0000-0001-6233-4311]{Andrew J. Rushby}
\affil{Department of Physics \& Astronomy, University of California, Irvine\\
4129 Frederick Reines Hall, Irvine, CA. 92697-4575}

\author[0000-0002-7086-9516]{Aomawa L. Shields}
\affil{Department of Physics \& Astronomy, University of California, Irvine\\
4129 Frederick Reines Hall, Irvine, CA. 92697-4575}

\author[0000-0002-2948-2811]{Manoj Joshi}
\affiliation{School of Environmental Sciences, University of East Anglia\\ 
Norwich, U.K., NR4 7TJ}



\begin{abstract}

The energy balance and climate of planets can be affected by the reflective properties of their land, ocean, and frozen surfaces. Here we investigate the effect of host star spectral energy distribution (SED) on the albedo of these surfaces using a one-dimensional (1-D) energy balance model (EBM). Incorporating spectra of M-, K-, G- and F-dwarf stars, we determined the effect of varying fractional and latitudinal distribution of land and ocean surfaces as a function of host star SED on the overall planetary albedo, climate, and ice-albedo feedback response. While noting that the spatial distribution of land masses on a given planet will have an effect on the overall planetary energy balance, we find that terrestrial planets with higher average land/ocean fractions are relatively cooler and have higher albedo regardless of star type. For Earth-like planets orbiting M-dwarf stars the increased absorption of water ice in the near-infrared (NIR), where M-dwarf stars emit much of their energy, resulted in warmer global mean surface temperatures, ice lines at higher latitudes, and increased climate stability as the ice-albedo feedback became negative at high land fractions.  Conversely, planets covered largely by ocean, and especially those orbiting bright stars, had a considerably different energy balance due to the contrast between the reflective land and the absorptive ocean surface, which in turn resulted in warmer average surface temperatures than land-covered planets and a stronger potential ice-albedo feedback. While dependent on the properties of individual planetary systems, our results place so constraints on a range of climate states of terrestrial exoplanets based on albedo and incident flux.

\end{abstract}
\keywords{atmospheres --- climate --- exoplanets --- ice}


\section{Introduction}

 The potential habitability of a given planetary environment -- that is the ability of that environment to support the activity of at least one organism (Cockell \textit{et al}., 2016) -- is strongly dependent on the amount of energy available in that system. To first order, planetary surface temperature provides the strongest control on the extent and distribution of habitable conditions, and is, in most cases, controlled by three main factors: the amount of incoming stellar radiation; the albedo or reflectivity of the surface on which that radiation is incident; and any potential greenhouse effect that may be caused by the absorption and remission of outgoing radiation by atmospheric gases such as carbon dioxide (CO$_2$) and water vapor (H$_2$O). 


Shields \textit{et al}. (2013), as well as other studies (Abe \textit{et al}., 2011; Joshi \& Haberle, 2012; Von Paris \textit{et al}., 2013) have demonstrated that bond albedo exhibits significant dependence on the stellar spectral type, particularly in the context of icy or frozen planetary surfaces. In this paper, unless otherwise stated, `albedo' refers to the \textit{Bond albedo} of a planet, which describes the total proportion or fraction of incident stellar flux, across all wavelengths, that is reflected back to space. This is in contrast to the geometric, spherical, or V-band albedo, which is wavelength and phase angle dependent. The relationship between albedo, temperature, and global ice cover represents a positive feedback within the planetary system, known as the ice-albedo feedback, that operates as ice cover advances equator-ward from the poles (due to a reduction in temperature) and planetary broadband albedo increases, thereby compounding the reduction in temperature and the further growth of icy surfaces of higher albedo (Lian \& Cess, 1977; von Paris \textit{et al}., 2013). Frozen surface features comprised of the same overlying material on planets around stars of different spectral types have considerably different Bond albedos, due to their spectral energy distributions (SED), which differ significantly. For example, smaller, redder stars, have peak output in the $\sim$0.8 to 1.2 $\mu$m range, where water ice and snow are particularly absorptive (Shields \textit{et al}., 2013; Joshi \& Haberle, 2012; von Paris \textit{et al}., 2013). The geometric albedo of ice and snow begins to decrease at wavelengths greater than $\sim$0.5 $\mu$m (see figure 1), and therefore the albedo of snow and ice covered surfaces on planets orbiting red dwarfs would be proportionally lower than that of the same surface on Earth (or any Earth-like planet in orbit around a F-, G-, or K-type star) (Joshi \& Haberle, 2012; Shields \textit{et al}., 2013). 

This unique radiative environment results in the Bond albedo of ice, for example, varying by $\sim$40\% between planets orbiting F-dwarf and M-dwarf stars, an effect which serves to dampen the ice-albedo feedback on M-dwarf planets relative to planets orbiting stars with greater output in the visible and near-UV (Shields \textit{et al}., 2013). The Bond albedos of land surfaces vary due to the SED of the host star, and also as a result of variations in their composition. Given the effect of albedo on the energy-balance of planets (Budyko, 1969; Sellers, 1969), the spectral dependence of surface albedo has a significant effect on the climate and long-term planetary habitability of terrestrial planets with land and ocean surfaces, depending on the spectral energy distribution of their host stars. Spectral type also has some control over the boundaries of the habitable zone (HZ), the effect of which is primarily evident at the inner, moist-greenhouse limit of the HZ, while the outer boundaries, at which sensitivity to the ice-albedo feedback is muted, seem to be relatively unaffected (Kasting \textit{et al}., 1993; Pierrehumbert, 2010; Shields \textit{et al}., 2013). However, the interaction between land and ocean fractional coverage, host star SED, and albedo in the context of planetary climate stability and ice-albedo feedback dynamics has not been previously explored. 

We are now aware of $>$4000 confirmed exoplanets, a small number of which are similar in size to the Earth (0.5 $>$ R$_{\oplus}$ $<$ 1.6), potentially rocky, and orbiting M-dwarf stars. Many of the surveys that have discovered exoplanets in the past (such as \textit{Kepler}), those that are currently under way (e.g. the Transiting Exoplanet Survey Satellite (\textit{TESS}) (Ricker, 2014)), as well as soon-to-launch (e.g. James Webb Space Telescope (\textit{JWST}) (Gardner \textit{et al}., 2006)) and future planned missions (e.g. the Large UV Optical Infrared Surveyor (\textit{LUVIOR}) (Kopparapu \textit{et al}., 2018)), focus on M-dwarf stars. 
Therefore, understanding the complex interactions between these planets and their host stars is crucial for determining and understanding their broader potential for long-term habitability, and for comparative planetology between similar terrestrial worlds in the orbit of stars of different spectral classes. However, we note that detailed assessments of the ability of a particular planet to support life are highly sensitive to the unique properties of the planetary and astrophysical environment (Cockell \textit{et al}., 2016).

In this work, we use a seasonally-varying, one-dimensional (1-D) (across latitude) energy balance model to investigate the relationship between ocean fractional coverage, broadband planetary albedo, and climate, as a function of host star SED. In subsequent sections, we will outline our methods and provide a description of the 1-D energy balance model that was adapted for use in this study. Our results are presented in $\S$3. $\S$4 contains a discussion of our findings, particularly in the context of planetary habitability, and outlines limitations and potential areas for future work.

\section{Methods and Models} \label{sec:methods}

Here, we describe our method of using a range of land, ocean, and icy/frozen surface types with empirically-derived albedos as input to a one-dimensional (1-D) across latitude, seasonally-varying energy-balance model (EBM). With the EBM, we explore the effect of land fraction on broadband planetary albedo as a function of the incident stellar SED.

After balancing incoming and radiated energy and accounting for area, the surface temperature of a planet can be approximated as:
\begin{equation}
{\sigma}T^4 = \frac{1}{4}(1-\bar{\alpha})L_{\odot}
\end{equation}

Where $\sigma$ denotes the Stefan-Boltzmann constant (5.67$\times$10$^{-8}$ W m$^{-2}$), \textit{T} is the surface temperature, $\bar{\alpha}$ denotes the planetary Bond albedo and $L_{\odot}$ is the incident flux from the star, normalized to the present-day Sun (1367 W m$^{-2}$) (Pierrehumbert, 2010). In this simple formulation, albedo takes the form of a single, zonally averaged value (Bond albedo) accounting for all wavelengths and phase angles. This final value is a normalized, weighted average of the mean `land albedo' ($\bar{\alpha_l}$) and the mean `ocean albedo' ($\bar{\alpha_o}$), which are in turn computed as a function of SED, land fraction, latitude ($\phi$) and temperature (\textit{T}).

The ice albedo ($\bar{\alpha_I}$) describes a representative surface comprised of a 1:1 mixture of course-grained ice and fine-grained snow, overlying either land or ocean surfaces, computed by Shields \textit{et al}. (2013), displayed in table 2 and figure 1. The initial land and ocean Bond albedos ($\bar{\alpha_{l_0}}$, $\bar{\alpha_{o_0}}$) are also taken from that source and vary by host star type. The stars used to compute SED are F-dwarf HD128167, K-dwarf HD22049 (Segura \textit{et al}., 2003), the G-dwarf the Sun (Chance \& Kurucz, 2010), and M-dwarf AD Leo\footnote{http://vpl.astro.washington.edu/spectra/stellar/mstar.htm \label{xx}}. These values were computed using a 1-D, multistream, multilevel, line-by-line, multiple-scattering radiative transfer model (Spectral Mapping Atmospheric Radiative Transfer Model (SMART)) (Meadows and Crisp, 1996) assuming Earth-like atmospheric conditions, including 64\% cloud coverage and Rayleigh scattering (Shields \textit{et al}., 2013). Instellation is spectrally integrated, and varies with both latitude and time of year in response to prescribed time-independent orbital parameters. The final albedo value for the land and ocean surfaces are then scaled proportionately to their fractional coverage, $F_L$, when used to compute surface temperature and average climate in the 1-D EBM.

\linespread{1.0}
\begin{table}[!htp]
\caption{Model Parameters used in this work} \label{tab:decimal}
\centering
\begin{tabular}{ll}
\hline
    \textbf{Parameter} & \textbf{Value/Range (units)} \\ \hline
    Incident Flux ($L_{\odot}$) & 1367 W m$^{-2}$ \\
    Eccentricity (\textit{e}) & 0.0167 \\
    Obliquity & 0 $^{\circ}$ \\
    Rotation Rate & 1 day(s) \\
    Land Fraction ($F_L$) & 0.01 -- 0.99 \\
    OLR when T = 273 K (\textit{A}) & 203.3 W m$^{-2}$\\
    OLR Temperature Sensitivity (\textit{b}) & 2.09 W m$^{-2}$ K$^{-1}$\\
    Heat Capacity, Land ($C_L$) & 0.45 W yr m$^{-2}$ K$^{-1}$ \\
    Heat Capacity, Water ($C_W$) & 9.8 W yr m$^{-2}$ K$^{-1}$ \\
    Diffusivity for Heat Transport (\textit{D}) & 0.44 W m$^{-2}$ K$^{-1}$\\
    Cloud fraction & 64\% \\
    Simulation Length & 30 yr \\
        \hline
    \end{tabular}
\end{table}

\subsection{1-D EBM} \label{subsec:EBM}

Energy-balance models describe a suite of models of varying complexity that are used to determine the energy-balance of a planetary body, and compute a possible climate solution. In this paper, we employ a modified version of the seasonally-varying 1-D EBM first described in North \& Coakley (1979) to describe the climate of the Earth, which calculates energy-balance at each latitude as the sum of absorbed shortwave radiation, outgoing longwave radiation (OLR), and the convergence of horizontal heat transport. This latter approximation is equated with vertical column heat capacity and mediated by a heat diffusion coefficient, which is set as proportional to the local meridional temperature gradient to produce an accurate representation of meridional heat transport. Roe and Baker (2010) have investigated the physical nature of the ice-albedo feedback parameterized in models of this sort, and note that the strength of the feedback depends linearly on the albedo contrast between the ice-covered and ice-free surfaces, while also being proportional to the efficiency of the heat distribution parameter (\textit{D}). For this work, rotation rate, obliquity, and eccentricity are assumed to be Earth-like, and incident flux for these simulations was held at the normalized present day value (1 \textit{S$_{\oplus}$} = 1367 W m$^{-2}$) (see Table 1). Land fraction (F$_L$) was varied between 0.01 and 0.99 distributed uniformly within each model latitude to represent a range of continent/ocean configurations to first-order. This method results in an average planetary land fraction that mirrors the uniform land fraction at each model latitude.  Sea ice is allowed to `form' when the temperature at a grid cell is below 271 K, which in turn alters the albedo to represent the overlying frozen surface. Ice accretion and ablation is determined by the relative difference between the top-of-the-ice heat flux and the ocean-ice heat flux, scaled by the ice density and remains independent of atmospheric water inventory (which cannot be set in the model), and therefore ice depth increases secularly with land fraction but we do apply a linear scaling with land fraction to the ice growth model in order to limit the thickness of ice sheets at high land fraction when atmospheric water vapor levels would be expected to be low. The ocean-ice heat flux is parameterized so that it is $\sim$4 W m$^{-2}$ at the lowest ice coverage (one grid cell), and then declines linearly to zero as ice covers the globe. Any heat supplied via the ocean-ice heat flux is subsequently `removed' from the ice-free ocean grid cells to ensure energy conservation. The ice temperature is maintained at 271 K by accretion or ablation based on energy conservation. Therefore, the variations in ice depth between planets orbiting different stars should be considered less as absolute depths of ice and more as relative values for comparison within the context of this model to identify areas/parameter space in which ice coverage is accumulating or receding. To compute average planetary climate, additional radiative forcing from radiatively active gases such as CO$_2$ and H$_2$O need also to be taken into account, represented here in the linearized form \textit{A} + \textit{bT}.  The coefficients \textit{A} and \textit{b} are taken from fits to satellite data from Earth (we use 203.3 W m$^{-2}$ and 2.09 W m$^{-2}$ K$^{-1}$, respectively, from North and Coakley (1979), but see also Cess (1976), North \textit{et al}. (1981) and Budyko (1969) for details on the derivation of these values) thereby constraining the applicability of this model to Earth-like atmospheres dominated by CO$_2$ and H$_2$O greenhouse gases.



\linespread{1.0}
\begin{table}[!htp]
\caption{Bond albedo values for various surface compositions used in these simulations.} \label{tab:decimal}
\centering
\begin{tabular}{ccccc}
\hline
     & \textbf{F-dwarf} & \textbf{G-dwarf} & \textbf{K-dwarf} & \textbf{M-dwarf} \\ \hline
    Land ($\bar{\alpha_l}$) & 0.414 & 0.415 & 0.401 & 0.331  \\ \hline
    Ocean ($\bar{\alpha_o}$) & 0.329 & 0.319 & 0.302 & 0.233 \\  \hline
    Ice ($\bar{\alpha_i}$) & 0.537 & 0.514 & 0.477 & 0.315 \\ \hline
        \hline
    \end{tabular}
    \tablecomments{These values were taken from table 1 in Shields \textit{et al}., (2013), and computed by a multiple-scattering radiative transfer code under conditions of an Earth-like atmospheric composition, cloud coverage, and Rayleigh scattering. `Ice' here refers to a 50$\%$ mix of large grained, blue marine ice and smaller grained snow particles that forms over both frozen land and frozen ocean cells.}
\end{table}

\begin{figure}[!htb]
\begin{center}
\includegraphics[width=0.85\linewidth]{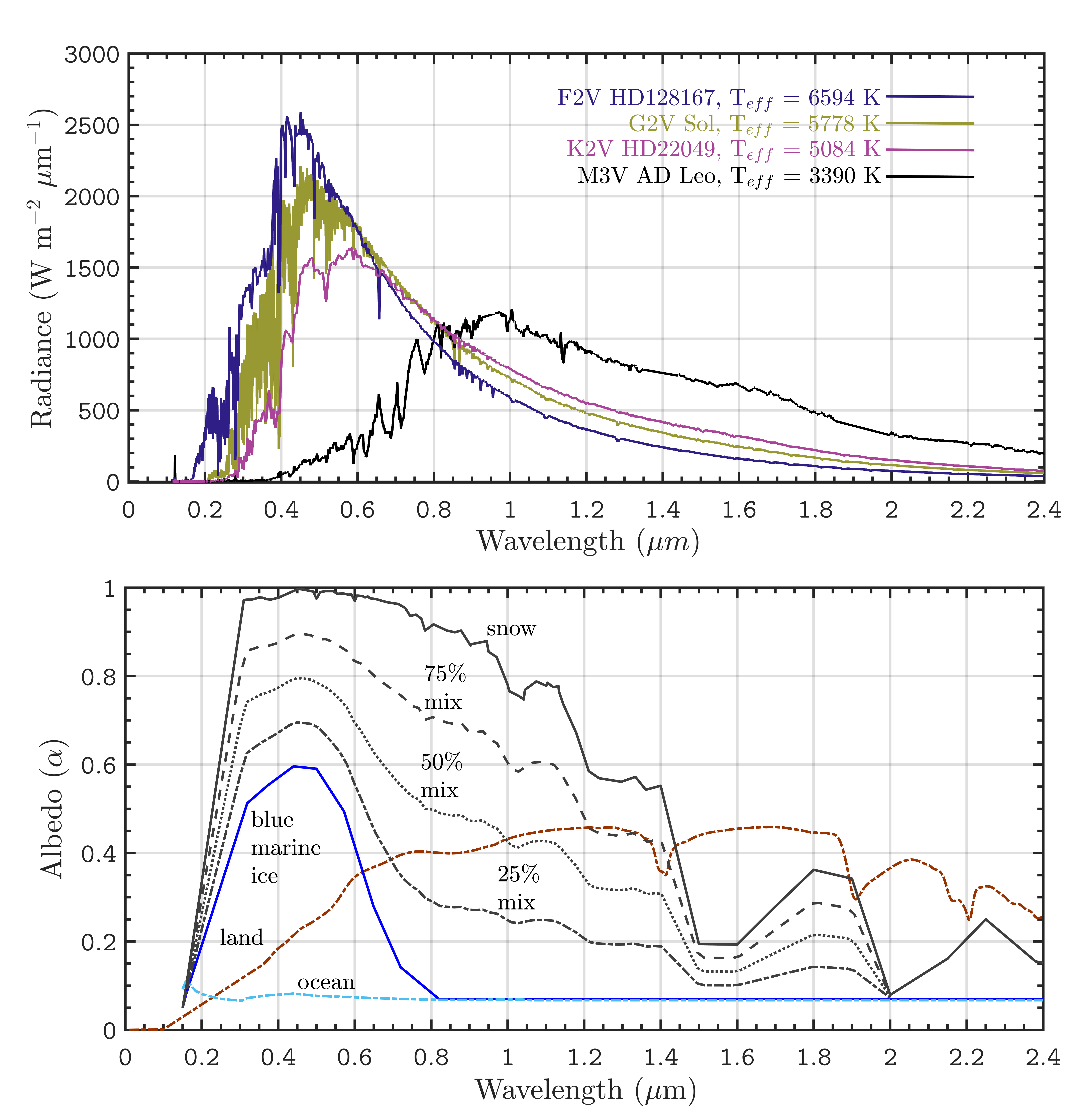}
\caption{Top: the SEDs for the F-, G-, K-, and M-dwarf stars used in this work. Below: The spectral distribution of the albedo of fine-grained snow, blue marine ice, and 25\%, 50\%, and 75\% mixtures of the two end-members. Ocean and land albedo spectral distributions are also shown (adapted from Shields \textit{et al}., 2013).}
\end{center}
\end{figure}

Table 2 lists the Bond albedos of the different surface compositions that are used in this paper, and Figure 1 their broadband reflectance spectra. The model uses the spectrally-independent albedo for the composition of the surface feature, surface temperature and latitude in order to recompute a final albedo and temperature. Convergence criteria were met when average surface temperature remained stable for 10 model years under no external forcings. Unfrozen land and water take the value of $\bar{\alpha_l}$ and $\bar{\alpha_o}$, respectively, and when frozen (i.e. \textit{T} $<$ $T_{i}$) both take the value of $\bar{\alpha_i}$, which in this paper refers to a 1:1 mix of fine-grained snow and course grained blue marine ice. The spectrum used for bare land in this work is that of the silicate clay mineral kaolinite taken from the USGS spectral library\footnote{https://speclab.cr.usgs.gov/spectral-lib.html}. We use kaolinite as a single representative land surface composite due to its relative commonality on Earth, and note that this spectrum has also been used in concert with this model in Shields \textit{et al}. (2013). Other workers, including Robinson \textit{et al}. (2011), who developed a whole-disk spectral model for Earth from EPOXI spacecraft observations, also use kaolinite as a representative composite, as does Meadows \textit{et al}. (2018) in their comparative planetology study of Proxima Centauri \textit{b}, in which they assume a 5.5\% kaolinite coverage on the modern Earth, or 23.1\% for early-Earth. 

\subsection{Model Validation} \label{subsec: ModelVal}

We validated the use of the EBM by reproducing Earth's bond albedo and average surface temperature at its present-day eccentricity, obliquity, approximate land/ocean ratio by latitudinal distribution (F$_L$ = 0.3), and incident stellar flux. Initial albedo values for the land, ocean, and ice-covered surfaces used in this validation simulation were taken from Shields \textit{et al}. (2013) using output from SMART for a Earth-like planet orbiting a G-dwarf star, and are outlined in more detail in the preceding paragraph.  We also explored the effect of differing compositions of icy surfaces, ranging from the extremely reflective, fine-grained endmember `snow' to darker, course grained `blue marine' ice, as well as intermediate mixtures (25\%, 50\%, 75\%) (see figure 1). Global surface temperatures varied by 0.7 K between the pure snow and blue ice cases, as the radiative effects of the change in albedos of these surfaces are confined to the high latitudes where instellation is low. For consistency with Shields \textit{et al}. (2013), we use the 50\% ice/snow mixture for our frozen land and ocean surfaces for the remainder of this work. The surface temperature returned by the model (284.1 K) was within 4 K (1.4\%) of the observed value (288 K). Albedo varies by planetary surface type, the distribution of which was set to represent an idealized case of the land and ocean distribution on the present Earth, and averaged 0.34 over the entire planet. This can be compared with the current broadband albedo for the Earth given by Stephens (2015) as 0.29. 


\subsection{Limitations and Future Work} 

The approach undertaken here is limited by the necessary parameterization of clouds, which are implicitly represented in albedo, and the climatic effect of a combined CO$_2$ and water vapor greenhouse, which is linearized, thereby restricting the applicability of our results to Earth-like atmospheric compositions. As we note in the previous section, ice growth in this model is temperature-dependent and occurs as a function of temperature contrast between the ocean cells and the ice-covered ocean cells, and not that of a dynamic water inventory. Sensitivity studies in which land and ice albedo was modified to simulate a lower rate of ice growth due to the lack of water at high land fractions revealed that planets around F- and G-dwarfs no longer experienced a global glaciations due to the lowered ice-land-ocean albedo contrast that increases climate stability and buffers the positive feedback response. We also expect these results to be affected by different land surface compositions, and expect that future modelling efforts to expand the scope of this work beyond Earth-like conditions using a 3D GCM. However, using a model of greater complexity has the drawback of increased computational burden and processor time. The approach we have used here as allowed us to explore a wide parameter space in a computationally efficient manner, while maintaining reasonable agreement with observations and other models.

\section{Results} \label{sec:results}

EBM simulations were carried out to investigate the effect of varying land fraction and star type on broadband planetary albedo, and mean global surface temperature. The results of these simulations are presented in figures \textit{2} -- \textit{4}. 

\begin{figure}[!htb]
\begin{center}
\includegraphics[width=\linewidth]{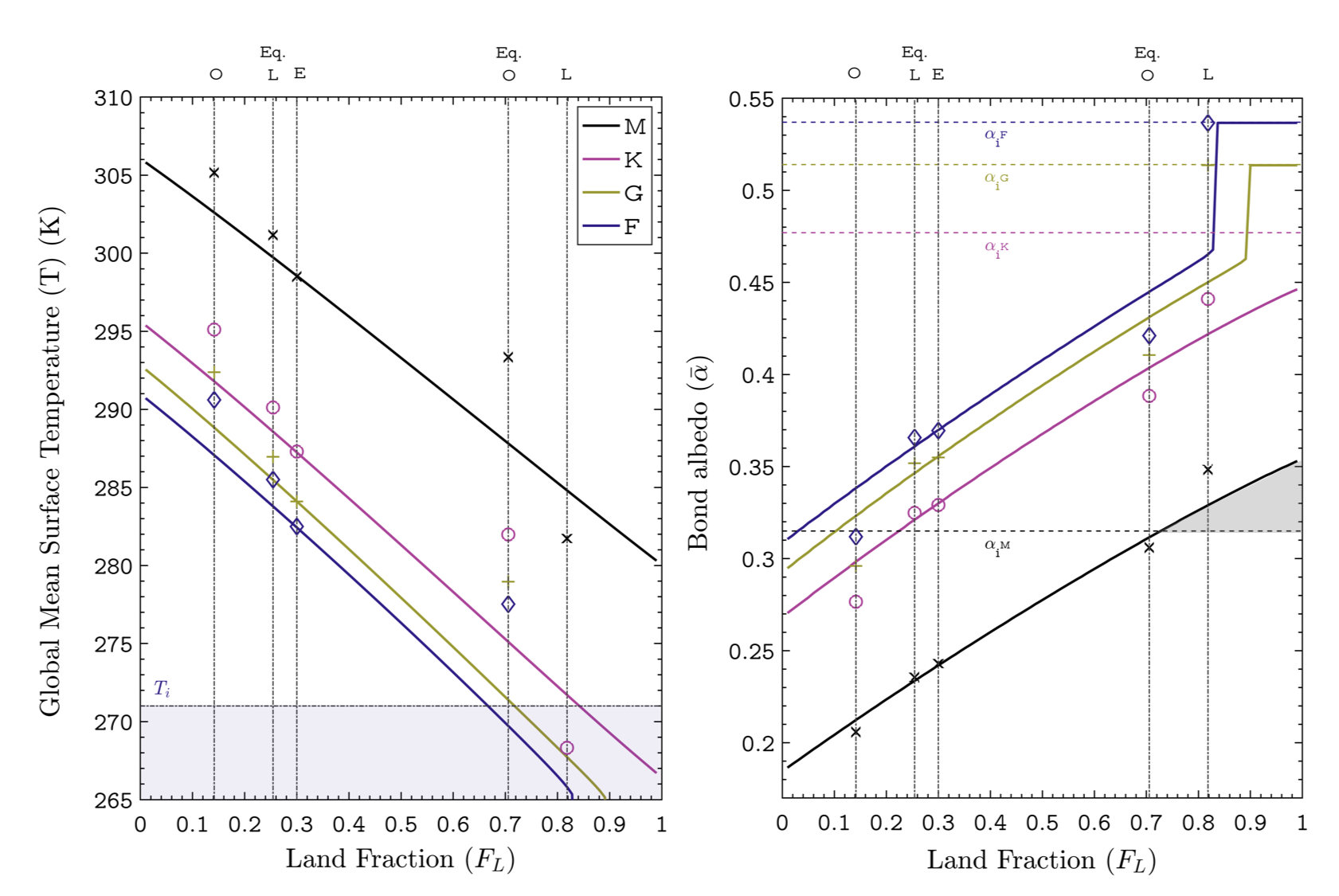}
\caption{Mean global surface temperature ($\bar{T}$) (left) and Bond albedo ($\bar{\alpha}$) (right) by average planetary land fraction ($F_L$) for terrestrial planets orbiting FGK and M-dwarf stars. The dashed lines labeled $\alpha_{i_{X}}$ denote the albedos of the 50\% ice-snow mixture on planets orbiting $X$-type star; the gray shaded area depicts a range of $F_L$ over which $\bar{\alpha}$ exceeds $\alpha_{i_{M}}$; the blue shaded region illustrates temperatures below 271 K. Markers indicate some individual land fraction and latitudinal distribution configurations tested in the model: present-day `Earth' (E), a scenario in which a narrow band of equatorial ocean exists on an otherwise land-covered planet (Eq. O), an `oceanplanet' (O) configuration with some land at the poles, an equatorial land belt on an otherwise oceanic planet (Eq. L.), and a `landplanet' (L) distribution with some open ocean at the poles. See text for discussion.}
\end{center}
\end{figure}

\begin{figure}[!htb]
\begin{center}
\includegraphics[width=\linewidth]{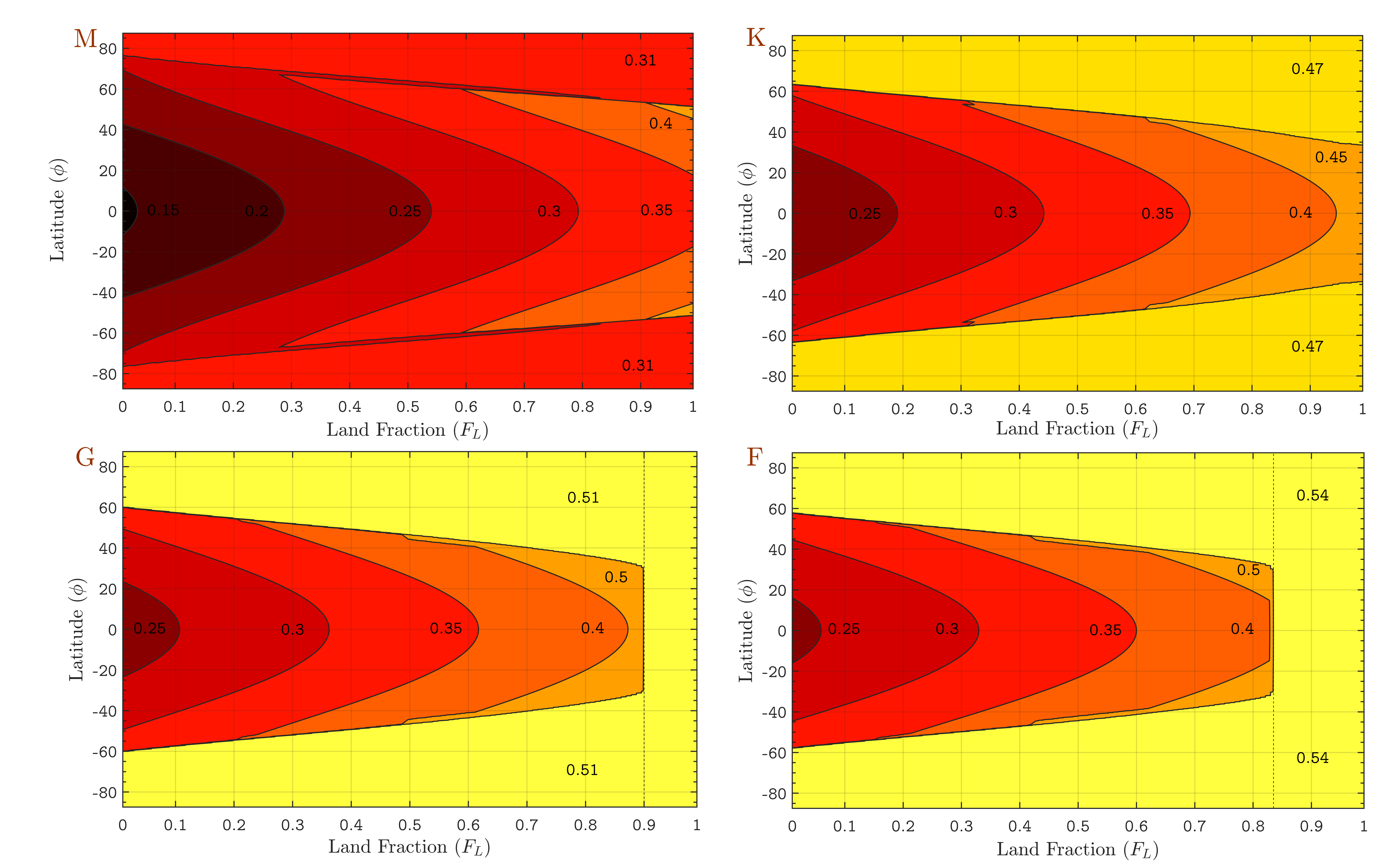}
\caption{Contours of albedo (dimensionless) as a function of land fractional coverage ($F_L$) and latitude ($\phi$) for a planet orbiting an M-dwarf (top left), K-dwarf (top right), G-dwarf (bottom left), and F-dwarf (bottom right) stars.}
\end{center}
\end{figure}

Ocean, land, and icy or frozen surfaces on planets orbiting M-dwarf stars have considerably lower Bond albedos than their analogues in the orbits of brighter stars, resulting in proportionately higher average global surface temperatures. The model output displayed in figures \textit{2} and \textit{3} illustrates surface albedo as a function of land fraction ($F_L$), latitude ($\phi$), and star type. Surfaces on planets orbiting F-dwarf stars have the highest albedos, and in general the albedo of a given planet increases with increasing land fraction. Figure \textit{2} also illustrates a region (shaded) in $F_L$ space ($F_L$ $>$ $\sim$0.72) at which the planetary Bond albedo is higher than the albedo of ice on M-dwarf hosted Earth-like planets ($\bar{\alpha}$ $>$ $\bar{\alpha_{i_{M}}}$) and the ice-albedo feedback becomes a negative, or stabilizing, feedback. We tested particular land distribution scenarios (i.e. configurations in which a non-uniform (by latitude) land fraction is used as input), including present-day Earth, and equatorial ocean and land belt configurations, that revealed some variation in surface temperature ($\sim$5 K) and albedo ($\sim$0.02), depending on the distribution of land in the model (see labelled markers in figure \textit{2}). In particular, ocean-dominated planets and land planets with oceanic equatorial belts (with otherwise identical ortibal parameters) have higher global mean surface temperatures when compared to a uniform land fraction distribution by latitude. Distinct zones of high and low albedo surfaces are shown in the simulations pertaining to planets orbiting F-, G-, and K-dwarfs (figure \textit{3}), which represent frozen (yellow) and unfrozen surfaces, respectively. Note that these frozen surfaces exhibit different Bond albedos depending on SED, despite being comprised of the same 50\% snow/ice mixture. Transitional boundaries between these zones are also evident, representing the maximum equator-ward extent of glacial conditions, i.e. a nominal `ice line' in latitude and F$_L$ parameter space, from which we can approximate the minimum land fraction required to force mid-latitude glaciations,assuming default model conditions: 0.46, 0.55, and 0.68 F$_L$ for planets around F-, G-, K-dwarfs respectively. 

M-dwarf planets do not exhibit mid-latitude glaciations in any of these simulations, although albedos for high land fraction planets in orbit around these stars are proportionally higher in the mid-latitudes and equatorial regions than at the frozen polar regions. A general trend exists in that planetary broadband albedo increases with increasing F$_L$ regardless of host star type, except in the case of the frozen polar regions where albedo remains constant in most cases due to the persistent frozen conditions at these latitudes.  Equatorial or low-latitude glaciations, analogous to a `snowball' or `slushball'-type event, are induced on Earth-like planets hosted by F-dwarf and G-dwarf stars when F$_L$ $>$ 0.8. For planets orbiting smaller, less luminous stars, equatorial glaciations are not induced by high land fractions, assuming an Earth-like atmospheric water vapor content, composition of the land surface, and the presence of variable clouds (see discussion).



\begin{figure}[!htb]
\begin{center}
\includegraphics[width=\linewidth]{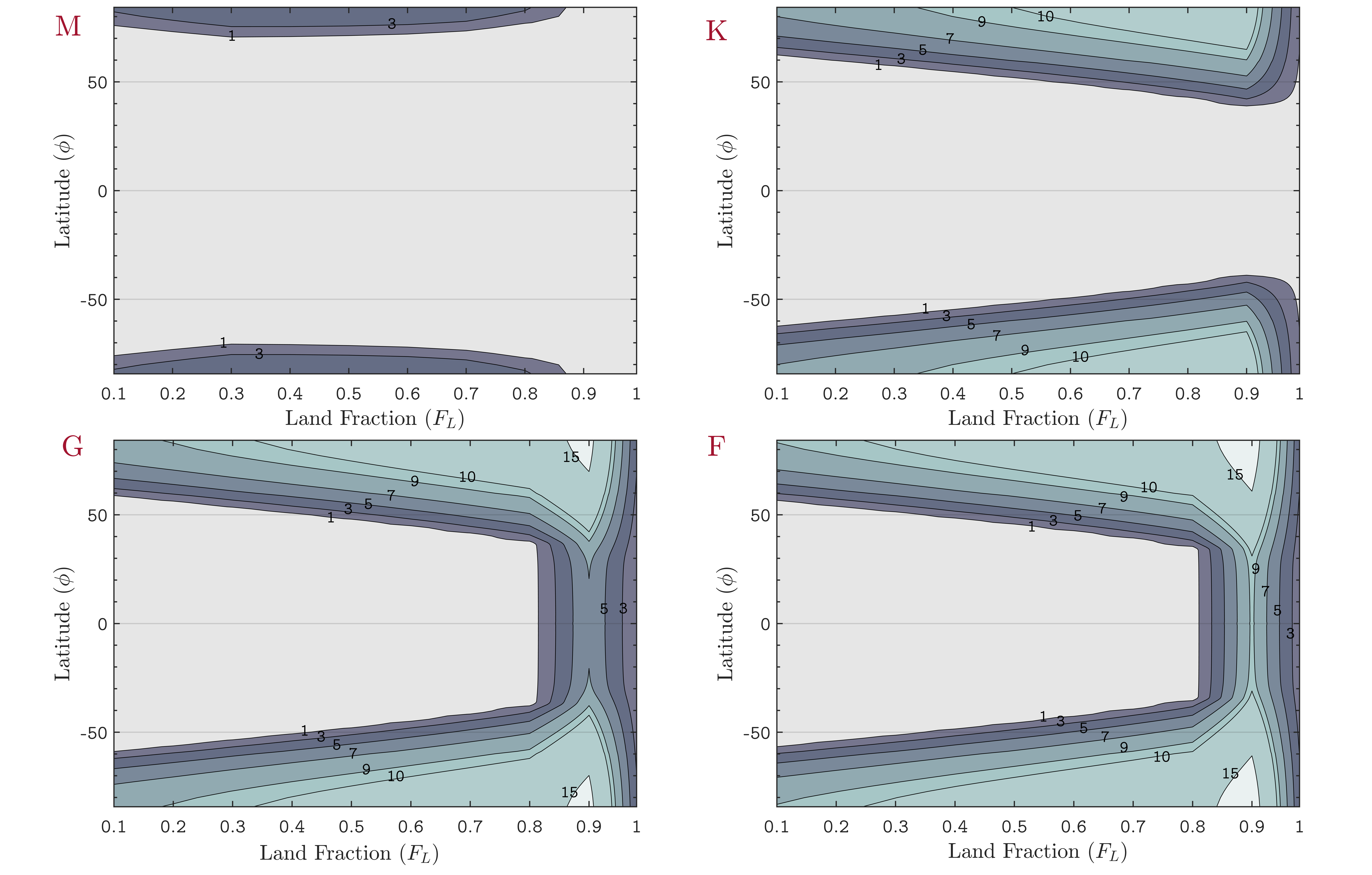}
\caption{Contours of approximate ice thickness ($>$1m) as a function of land fraction ($F_L$) and latitude ($\phi$) for planets orbiting M (top left), K (top right), G (bottom left), and F-type (bottom right) stars.}
\end{center}
\end{figure}

Host star SED and land fraction also influences the maximum equatorward extent of persistent frozen surface conditions, taken here to refer to areas in which ice of at least 1 meter in thickness persists for the 30 model-year simulation (Figure \textit{4}). This figure displays contours of approximate ice thickness by land fraction and latitude, and demonstrates that planets orbiting F- and G- dwarf stars are especially prone to persistent and substantial ice formation extending to the mid-latitudes and equatorial regions beyond the $F_L$ glaciation limits discussed earlier. Additionally, these planets exhibit frozen surface conditions between 15 and 20 degrees further towards the equator, albeit with uniformly distributed land fraction over latitude, than in the case of an M-dwarf planet. This comparison suggests a lower sensitivity towards the ice-albedo feedback mechanism and more stable long-term climate states on M-dwarf land planets, other conditions remaining the same. As ice growth in this model is driven by temperature, we note a peak in ice depth at high land fractions ($\sim$0.9) when temperatures are lowest for planets orbiting G- and F-stars. This is followed by a reduction in thickness as $F_L$ approaches unity due to the scaling parameterization discussed earlier in this work that prevents unrealistically thick ice sheets forming on dry, land-dominated worlds. 

Planets with a greater ocean fraction, under otherwise identical conditions, absorb considerably more incident energy and are on average warmer as a result (see Figures \textit{2}-\textit{4}). When considering a planet orbiting a G-dwarf host (with otherwise `Earth-like' properties), increasing the fraction of the surface covered by continent reduces the global average surface temperature by 10 - 15\% between simulations using 10\% land coverage ($\bar{{T}_{s}}$ = 290 K) and those using 90\% land surface coverage ($\bar{{T}_{s}}$ = 258.3 K) (see Figure \textit{2}). While this relationship holds for all star types considered in this work, Earth-like planets orbiting M-dwarf stars remain above the threshold for freezing conditions even when covered predominately by land, maintaining an average surface temperature of $\sim$303.9 K at 10\% and 282.6 K at 90\% land fractional coverage, respectively. An average a M-dwarf `aquaplanet' (F$_L$ $<$ 0.15) remains $\sim$15 K warmer than the same type of planet around an F-dwarf when both are receiving the equivalent stellar flux as the present Earth.



\section{Discussion} \label{sec:discuss}

The results presented in this paper suggest that planets dominated by land are, on average, cooler and more highly reflective than those with higher fractional ocean coverage. We note a sharp transition between partially and fully glaciated planets, forced by the ice-albedo feedback, that represents the land fraction threshold at which ice growth extends from the poles to the latitudinal `tipping point' for global glaciation. Once ice growth extends beyond this limit, the ice-albedo feedback enters a runaway state and the transition to globally glaciated conditions is sharp and rapid; this threshold varies with star type as the surface albedo contrast is wavelength dependent. The magnitude of the runaway response is sensitive to several parameters, including heat distribution efficiency and land/ice/ocean albedo contrast, which is controlled by land fraction, as well as stellar flux and SED.  

Given the lower albedo and increase in absorbed incident flux, coupled with the findings of previous work in this area (Shields \textit{et al}., 2013; Joshi \& Haberle, 2012), we would expect a significant climatological effect due stellar spectral energy distribution affecting the albedo of equivalent surfaces on planets orbiting M-dwarf stars compared to those around larger K-, G-, and F-dwarf stars. We find that, due to the reduction in the reflectance of water ice at wavelengths longer than $\sim$1$\mu$m, planets orbiting M-type stars are warmer and represent the only case in this model formulation where non-glacial conditions can exist on a largely land-covered planet. This effect is also pronounced at higher latitudes, where surfaces that would freeze on planets around more luminous stars remain unfrozen on M-dwarf planets, resulting in overall warmer global mean surface temperatures. Due to the interaction between the unique spectral energy distribution of M-dwarf stars and the optical properties of frozen surfaces on orbiting planets, the reversal of the effect of the ice-albedo feedback results in greater shortwave absorption and ultimately a stabilization of the climate in response to further cooling. The relative equator-ward extent of the ice line is also less evident. This effect is especially pronounced at low land fractions, and suggests that above $\sim$0.85 F$_L$, the surfaces of M-dwarf planets may remain unfrozen even at polar latitudes at Earth-equivalent incident flux. Seasonal variations in temperature are also dampened for planets dominated by ocean, as land has a lower specific heat capacity that modulates the seasonal cycle of temperature. 

Idealized M-dwarf `aquaplanets' are also on average $\sim$15 K warmer than an equivalent ocean planet orbiting a G-dwarf. While ocean albedo remains relatively similar across the visible and infrared wavelengths, the strong reflectance of marine and terrestrial icy surfaces in the shorter wavelengths at which the F-, G-, and K- stars experience their peak output (0.4 - 0.7 $\mu$m) dominates the radiative balance of these planets. This serves to reduce the amount of absorbed stellar flux at the surface and results in the formation of highly reflective frozen areas that are resilient to changes in radiative forcing by reducing both the amount and pole-ward extent of darker, unfrozen ocean areas. 

Our results vary somewhat from Abe \textit{et al}. (2011), who used a 3D-GCM with a dynamic hydrological cycle to investigate changes to the boundaries of the habitable zone for dry and moist planets orbiting G-dwarfs. Abe \textit{et al}. (2011) found land planets to be more difficult to freeze than their aquaplanet counterparts, due to unsaturated atmospheric conditions above the tropics facilitating greater longwave emission, lower rates of hydrogen escape, and lower thermal inertia. However, this work focussed on G-dwarf host stars in their simulations and a dependence of snow and ice deposition on atmospheric moisture, the albedo of which was not wavelength-dependent and scaled linearly with temperature between 0.5 and 0.75. In contrast, Turbet \textit{et al}. (2016) found that when using a GCM to explore a range of potential volatile inventories for the terrestrial exoplanet orbiting the nearby mid-type M-dwarf star Proxima Centauri, a drier atmosphere and planet surface resulted in cooler climates. This difference is due to a reduction in the water vapor greenhouse effect on dry, land-covered planets. While the effect was not taken into account in this work, the increased IR emission from M-dwarfs relative to G-dwarf stars would likely result greater greenhouse warming from a water vapour atmosphere at equivalent solar flux distances thereby further accentuating the effect of the ice-land albedo contrast seen here (Kasting \textit{et al}., 1993; Shields \textit{et al}., 2013). At low land fractions (aquaplanets) the effect would be considerable, while land-dominated planets around M-dwarfs would remain warmer with higher latitude ice lines.  Therefore, it is likely that further investigations of this effect over a range of land fractions and distributions and star types will be necessary to better determine the complex response of the climate system under these conditions.

Our model assumed an atmospheric water content (in the context of a greenhouse forcing) parameterized for an Earth-like land/ocean fraction and distribution. We performed sensitivity studies by scaling the OLR temperature sensitivity parameter with land fraction (assuming a G-dwarf planet with the otherwise default conditions outlined in table 1), and found the results to vary little from our original results at low to moderate land fractions. At high land fractions (F$_L$ $>$ 0.7) this scaling results in planets that are considerably cooler (-20 K) than we found with our default parameterization, but this simplified approach is likely pushing the OLR linearization beyond its intended range of applicability. We also tested OLR linearization parameterizations from Budyko (1969), Cess (1976), and Lindzen and Farrell (1977). While all of these parameterization sees a marked decline in global mean temperatures as land fractions approach unity, the primary variation between these schemes occurs at high land fractions where the latter two fits diverge to be very much cooler than our default case. Increases in \textit{b} result in a greater climate sensitivity (i.e. a larger temperature response to a given perturbation). Our results therefore constitute a conservative lower limit on the possible differences in climates between planets with high and low land fractions. A more complex treatment of atmospheric water vapor requires a dynamic hydrological cycle and a coupled ocean-atmosphere, which is beyond the scope of this work.

We also assumed that the average land fraction used here is representative of a planet with uniformly distributed land by latitude, but recognize that heterogeneous continental distributions (for example, the `aquaplanet', `equatorial ocean', `equatorial land', or a `landplanet' configurations shown in Figure \textit{2}) that exhibit the same average land fraction will result in a somewhat different energy balance from these idealized cases depending on the specific configuration of land and ocean due to the difference in specific heat capacities and thermal diffusivity between these two surface types. Furthermore, other land surface composites and mixtures have been considered in the context of exoplanet studies (e.g. Hu \textit{et al}., 2012), and we expect that using a composite with higher Bond albedo will proportionately increase the land-ocean-ice contrast and therefore lower the surface temperatures of land dominated worlds, while also reducing climatic stability; a lower albedo land composition will lower this contrast and thereby buffer climate stability to a greater degree than our default composite. 

F-dwarf planets remain consistently cooler and more reflective than analogues orbiting smaller stars, and figures \textit{2} through \textit{4} illustrate that planets around these stars could experience equatorial glaciation at high land fractions ($>$0.8) even when receiving the equivalent incident flux as present day Earth. The high output of stars in this class in the 0.4-0.5 $\mu$m wavelength range, where ices are highly reflective, contributes to the proportionately higher albedo of these surfaces. This increased surface reflectivity and cooling also affects the location of the `ice line' (the minimum latitude at which glacial conditions are present), which extends relatively further towards the equator in the case of brighter stars; for an F-dwarf planet with land fraction coverage of 0.2 F$_L$ the ice line falls at approximately 52 degrees, whereas the same fractional coverage and incident flux around an M-dwarf sets the ice line closer to 70 degrees (see figure \textit{4}). 

The potential for lower latitude glaciation also affects the stability, and feedback strength, of the ice-albedo feedback mechanism itself. Joshi \& Haberle (2012) note that changes to the ice albedo affect the `ice-albedo response' term (\textit{I}), a component of the `clear-sky feedback parameter' (along with the black body response (\textit{B}) and the water vapor feedback (\textit{W})). This approach can be used to approximate the temperature response (\textit{dT}) of a radiative forcing (\textit{dF}) by \textit{dT} = \textit{dF} / (\textit{C} + \textit{B} + \textit{W} + \textit{I}), where \textit{C} is the cloud feedback. These feedback parameters are in units of W m$^{-2}$ K$^{-1}$, and will be negative if the feedback is positive (e.g. \textit{I} for Earth $\approx$ -0.3 W m$^{-2}$ K$^{-1}$). Decreasing albedo, in the case of icy surfaces around M-dwarfs, increases \textit{I} thereby reducing the total climate response to a change in radiative forcing, whereas an increase in albedo associated with F-dwarf hosts would decrease \textit{I} and increase the planetary climate response to a given perturbation. The results presented in this work suggest that \textit{I} reverses sign and becomes positive (i.e. a negative or stabilizing feedback) when considering M-dwarf planets with high land surface coverage. However, the precise magnitude of this response is contingent on the properties of the individual planetary system under consideration, and would be sensitive to variations in atmospheric composition, planet size, incident flux, as well as long-term orbital effects on climate (such as eccentricity, obliquity, precession) (Deitrick \textit{et al}., 2017, 2018). Additionally, the uniform distribution of land fraction with latitude, though idealized, allows for the normalization of topographic variability across planets for the purposes of comparative planetology.

\subsection{Habitability and Observability} 

The modelled differences in the ice-albedo response on planets around main-sequence stars of different spectral types have considerable implications for habitability and observability. The increasing climatic stability exhibited by the modelled M-dwarf planets (assuming otherwise Earth-like atmospheric conditions) regardless of land fraction suggests a boon to the habitablity potential of these worlds, while the contrasting trend (lower climate stability, lower surface temperatures for a given incident flux, and lower latitude ice lines) for F-dwarf planets may make these worlds less amenable to life. However, we note that the effects a negative ice-albedo feedback on plnaetary habitability are complex and uncertain, given that Earth remains habitable while exhibiting a positive ice-albedo feedback, and further study of this phenomena is warranted. Previous work finds an anti-correlation between stellar mass and planet occurrence rate, noting that M-dwarfs host 3.5 times more planets in the 1 - 2.8 Earth radii ($R_{\oplus}$) range than main-sequence FGK stars, but half the number of planets larger than 2.8 $R_{\oplus}$ (Mulders \textit{et al}., 2015; Shields \textit{et al}., 2016). These systems are also more likely to host a transiting planet in the HZ (Charbonneau \& Deming, 2007). Additionally, in terms of observational prospects, the near-term detection of Earth-sized and super-Earth planets ($>$2 $R_{\oplus}$) from TESS, as well as follow-up studies carried out by ground-based radial velocity instruments or JWST, are likely to focus on M-dwarf systems (Sullivan \textit{et al}., 2015; Barclay \textit{et al}., 2018).  The direct measurement of planetary albedo, which has been the primary focus of this work, is possible in the case of transiting giant planets using secondary eclipse measurements, but this technique is unlikely to be applicable to smaller, rocky planets due to the high transit depth and signal-to-noise required (Seager \& Deming, 2016). However, the transmission spectrum of the super-Earth planet GJ 1214\textit{b} provided the first evidence of a cloudy or hazy terrestrial exoplanet atmosphere (Kriedberg \textit{et al}., 2014), and future space-based observatories such as JWST, as well as large ($>$30 m) ground-based telescopes, may provide further spectroscopic characterization of major molecules in the atmospheres of M-dwarf super-Earths for climate and habitability assessments (Seager \& Deming, 2016). Kriedberg \& Loeb (2016), Turbet \textit{et al}. (2016) and Wolf \textit{et al}. (2019) find that detecting and characterizing the potential atmosphere of nearby, Earth-like exoplanet Proxima Centauri b could be achieved by JWST using thermal phase curve observations to detect variations in heat distribution of a planet with an atmosphere versus one comprised of bare-rock. The results presented here suggest a general trend between land fraction, albedo, and climate based on star type and incident flux, which is valuable for determining the reflectivity, emmissivity and overall energy-balance of the surface. However, future observations of terrestrial planets will be affected by the composition of their atmospheres, as well as the presence of clouds and/or hazes, and, specifically in the case of planets around M-dwarf planets, UV-driven water dissociation and hydrogen escape (Luger \& Barnes, 2015). Given their relative commonality (comprising $\sim$70\% of the main-sequence stars in the stellar neighborhood), high small planet occurrence rate, and future observational opportunities, considerable theoretical and modelling efforts remain focused on the climates and potential habitability of M-dwarf systems (Shields \textit{et al}., 2016). The effects of land and ocean distribution on long-term climate stability presented here should be considered in habitab ility metrics of terrestrial planets discovered around low mass stars.

\section{Conclusions} \label{sec:conc}

Using a 1-D energy balance model incorporating wavelength-dependent albedos for a range of surface types (land, ocean, water ice and snow), we quantified the effect of land and ocean fractional surface coverage on the broadband albedo and climate of Earth-like planets orbiting M-, K-, G-, and F-dwarf stars. Planets with a higher fractional coverage of land are cooler and more reflective regardless of star type, thereby increasing the sensitivity and response of the ice-albedo feedback on heavily land-dominated planets. Coupled with the lower Bond albedo of ice, land, and open ocean on planets hosted by M-dwarf stars, these planets, uniquely, can exhibit relatively higher fractional land coverage and remain unglaciated as the ice-albedo feedback reverses sign and becomes stabilizing. Ices on planets hosted by large, hot stars have a higher Bond albedo (due to the high output of these stars at shorter wavelengths) and therefore exhibit a greater sensitivity to changes in radiative forcing, which includes variations in land coverage; high land fractions result in a strong ice-albedo feedback and global glaciations beyond $F_L$ $>$ $\sim$0.8 and $F_L$ $>$ $\sim$0.89 for Earth-like planets hosted by F- and G-dwarf stars, respectively. `Aquaplanets' (those dominated by ocean) are consistently warmer on average than those with greater land fractions, and exhibit lower seasonal variation in temperature due to the higher heat capacity of water. This effect is particularly pronounced on planets orbiting M-dwarfs, which may be $\sim$15 K warmer than a similar world orbiting an F-dwarf star. Earth-like planets hosted by M-dwarf stars are able to maintain a global mean surface temperature of 283 K at 0.9 $F_L$ and 304 K at 10\% land fraction.


\section{Acknowledgements}
This material is based upon work supported by the National Science Foundation under Award No. 1753373, and by NASA under grant number NNH16ZDA001N, which is part of the NASA ``Habitable Worlds" program. We thank Cecilia Bitz and Igor Palubski for helpful discussions regarding this work, and Eric Wolf for providing the spectrum for AD Leo for figure \textit{2}.



\end{document}